\documentclass[12pt]{iopart}

\usepackage{iopams}  
\usepackage{graphicx}
\begin{document}

\title[An optimization of the screw pitch]{
$d=2$ 
transverse-field Ising model
under the
screw-boundary condition: 
An optimization of the screw pitch
}

\author{Yoshihiro Nishiyama}

\address{
Department of Physics, Faculty of Science,
Okayama University, Okayama 700-8530, Japan
}

\begin{abstract}
A length-$N$ spin chain with
the $\sqrt{N}(=v)$-th neighbor interaction is
identical to a 
two-dimensional ($d=2$) model under the screw-boundary (SB) condition.
The SB condition provides a flexible scheme to construct a 
$d \ge 2$
cluster from an arbitrary number of spins;
the numerical diagonalization combined with the SB condition
admits a potential applicability to a class of systems
intractable with the
quantum Monte Carlo method due to the negative-sign problem.
However, the simulation results suffer from 
characteristic finite-size corrections inherent in SB.
In order to suppress these corrections,
we adjust the screw pitch $v(N)$ so as to minimize the excitation gap
for each $N$.
This idea is adapted to
the
transverse-field Ising model 
on the triangular lattice 
with $N \le 32$ spins.
As a demonstration, the 
correlation-length critical exponent $\nu$
 is analyzed 
rather in detail.
\end{abstract}

\maketitle

\section{\label{section1}
Introduction}

A length-$N$ spin chain with the
$\sqrt{N}(=v)$-th neighbor interaction is identical to 
a two-dimensional 
($d=2$)
model under the screw-boundary condition;
a schematic drawing is presented in Fig. \ref{figure1}.
The screw-boundary condition provides a flexible scheme to 
construct a 
$d \ge 2$
cluster from an arbitrary number of spins $N$.
With the aid of Novotny's method
\cite{Novotny90,Novotny92},
one is able to implement 
the screw-boundary condition systematically;
details are explicated afterward.
Because the quantum Monte Carlo method is inapplicable to
a class of systems such as
the frustrated quantum magnets (negative-sign problem),
the numerical diagonalization combined with Novotny's method
might provide valuable information 
as to such an intriguing subject.

As a matter of fact,
the screw-boundary condition was adapted to
the $S=1$ triangular  magnet
for $N \le 20$
\cite{Nishiyama11};
tuning the 
spatial anisotropy and the biquadratic interaction,
we are able to realize 
the deconfined criticality
\cite{Singh10}
between 
the resonating-valence-bond (RVB)
and magnetic phases.
The present scheme (modification) is readily applicable to this problem.
Moreover,
a recent reexamination of the transverse-field Ising model
in $d=2$ and $3$ via the real-space renormalization group
revealed that a procedure, the 
so-called ``exact invariance under renormalization,'' yields 
a substantially
improved
critical exponent even for an analytically tractable system size;
a relevance to the present treatment is addressed in the last section.

In this paper,
we make an attempt to reduce the finite-size corrections inherent in the 
screw-boundary condition.
As a matter of fact,
the simulation results exhibit 
bumpy
finite-size deviations
depending on the condition whether
the pitch $v(=\sqrt{N})$ is close to an 
integral
number or not; actually,
in Ref. \cite{Novotny90}, 
such a singular case is treated separately.
The screw pitch $v$ admits a continuous variation,
and there is no reason to stick to a fixed pitch $v=\sqrt{N}$. 
We adjust $v(\approx \sqrt{N})$ so as to 
minimize 
the excitation gap $\Delta E$ 
for each system size $N$.
As a demonstration, we adapt this idea to
the 
(ferromagnetic)
transverse-field Ising model on the triangular lattice
for $N \le 32$.
In order to demonstrate the performance of this approach,
we analyze the 
criticality (correlation-length critical exponent $\nu$)
rather in detail;
the simulation results are compared with those of
the fixed screw pitch $v=\sqrt{N}$.

The rest of this paper is organized as follows.
In Section \ref{section2},
we explicate the simulation algorithm,
placing an emphasis on the scheme how we fix the screw pitch
to an optimal value.
In Section \ref{section3},
we show the numerical results.
As a demonstration, we make an analysis of the 
critical exponent $\nu$.
In Section \ref{section4},
we present the summary and discussions.

\section{\label{section2}
Simulation algorithm for 
the two-dimensional transverse-field Ising model
under
the screw-boundary condition:
Novotny's method}

We employ 
Novotny's method in order to implement the screw-boundary condition
\cite{Novotny90,Novotny92}.
For the sake of selfconsistency, 
in this section,
we present an explicit simulation algorithm.

Before commencing an explanation of the technical details,
we sketch a basic idea of Novotny's method; see Fig. \ref{figure1}.
We implement the screw-boundary condition for a finite cluster
with $N$ spins.
We place an $S=1/2$ spin (Pauli matrix $\vec{\sigma}_i$)
 at each lattice point 
$i(\le N)$.
Basically, the spins 
 constitute a one-dimensional ($d=1$) structure.
The dimensionality is lifted to $d=2$ by the long-range interactions
over the $\sqrt{N}(=v)$-th-neighbor distances;
owing to the long-range interaction, the $N$ spins form a
$\sqrt{N}\times \sqrt{N}$
rectangular network effectively.
(The significant point is that the 
screw pitch $v \approx \sqrt{N}$ is not necessarily
an integral number, and can even afford to vary continuously.)

According to Novotny
\cite{Novotny90,Novotny92}, the long-range interactions
are
introduced systematically
by the use of the
translation operator $P$; see Eq. (\ref{TP_decomposition}).
The operator $P$ satisfies the formula
\begin{equation}
P | \sigma_1,\sigma_2,\dots,\sigma_N \rangle
   = | \sigma_N,\sigma_1,\dots,\sigma_{N-1}\rangle  .
\end{equation}
Here, 
the Hilbert-space bases 
$\{| \sigma_1,\sigma_2,\dots,\sigma_N \rangle \}$
($\sigma_i = \pm 1$) 
diagonalize the operator $\sigma^z_i$;
\begin{equation}
\sigma_j^z | \{ \sigma_i \} \rangle  =
\sigma_j   | \{ \sigma_i \} \rangle  .
\end{equation}

\subsection{Details of Novotny's method}

We formulate the above idea explicitly.
We adapt Novotny's method to the transverse-field Ising model
on the triangular lattice.
The Hamiltonian with a screw pitch $v(\approx \sqrt{N})$ 
is given by
\begin{equation}
\label{Hamiltonian}
{\cal H}
= -J \left[
H \left(1\right)+
H \left(v+\frac{1}{2}\right)+
H \left(v-\frac{1}{2}\right)
\right]
-\Gamma \sum_{i=1}^{N} \sigma^x_i
,
\end{equation}
with the transverse magnetic field $\Gamma$.
Hereafter, we consider the exchange interaction
$J$ as a unit of energy ($J=1$).
The 
$v$-th neighbor interaction $H(v)$ 
is a diagonal matrix, whose diagonal element is given by
\begin{equation}
\label{TP_decomposition}
H_{ \{\sigma_i\},\{\sigma_i\} }(v)
=\langle \{\sigma_i\} | H(v) | \{\sigma_i\} \rangle
=\langle \{\sigma_i\} | TP^v | \{\sigma_i\} \rangle
   .
\end{equation}
The insertion of $P^v$ is a key ingredient to introduce
the $v$-th neighbor interaction.
Here, the matrix $T$ denotes
the exchange interaction between
$\{ \sigma_i \}$ and $\{ \tau_i \}$;
\begin{equation}
\label{plaquette_interaction}
\langle \{\sigma_i\} |T| \{\tau_i\} \rangle=
\sum_{k=1}^{N}
\sigma_{k}\tau_{k}  
  .
\end{equation}
Afterward, we search for an optimal value of
the 
screw pitch $v(\approx \sqrt{N}) $.

The above formulae complete the formal basis of our simulation
scheme.
We diagonalize the Hamiltonian matrix
(\ref{Hamiltonian})
for $N \le 32$ spins numerically.
In the practical numerical calculation,
however,
a number of formulas may be of use;
see
the Appendices of
Refs. \cite{Nishiyama07b,Nishiyama10}.

\subsection{Search for an optimal screw pitch $v$}

An optimal value of the screw pitch
$v$ is determined as follows.
In Fig. \ref{figure2},
we plot the (first) excitation gap $\Delta E$ for the parameter 
$a=v^2/N$,
and the system size, $N=$ ($+$) $28$ and ($\times$) $32$.
Here, 
we fix
the transverse magnetic field to $\Gamma=4.8$,
which is close to the critical point
(separating the paramagnetic and ferromagnetic phases); see Fig. \ref{figure3}.
(An underlying idea behind this analysis is presented afterward.)
From Fig. \ref{figure2},
we see that
the excitation gap takes a minimum value beside $a=1$;
note that so far, the screw-boundary-condition
(Novotny method) simulation has been performed at 
$a=1$ ($v=\sqrt{N}$).
We adjust $v(N)$ so as to minimize the excitation gap for each $N$.
As a result,
we arrive at an optimal screw pitch ($v=\sqrt{Na}$)
\begin{equation}
\label{adjusted_v}
a=
1.12 \ ,\   
0.98 \ ,\  
0.81 \ ,\  
0.68 \ ,\  
1.30 \ ,\  
{\rm and} \ 
1.06 \ ,\  
\end{equation}
for $N=12,16,\dots,32$, respectively.
(The series of results may indicate a regularity such that 
$\sqrt{N}$-dependence of $a$ is like
 sawtooth.)
The optimal screw pitch differs from the conventional
one \cite{Novotny90,Novotny92}
\begin{equation}
\label{fixed_v}
v=\sqrt{N} .
\end{equation}
A comparison between Eqs. (\ref{adjusted_v})
and (\ref{fixed_v}) will be made in order to elucidate the performance
of the present approach.

The underlying idea behind the above screw-pitch optimization 
(\ref{adjusted_v})
is as follows.
In general,
the excitation gap 
is proportional to reciprocal correlation length $\xi \propto 1/\Delta E$.
On the one hand, 
according to the scaling hypothesis,
namely, at the critical point,
the correlation length develops up to the linear dimension 
of the cluster $L \sim \xi$; 
in other words,
the correlation length is bounded by the system size,
and it should extend fully for the 
equilateral cluster.
Hence, 
as the energy gap gets minimized,
the (effective)
shape of the cluster restores its spatial isotropy,
rendering the finite-size behavior smoother.

\section{\label{section3}
Numerical results}

Setting the screw pitch to an optimal value (\ref{adjusted_v}),
we turn to the analysis of the exponent $\nu$ rather in detail.
In Fig. \ref{figure3}, we plot
the scaled excitation gap for various $\Gamma$ and $N=12,16,\dots,32$;
the system size $L$ is given by $L=\sqrt{N}$, because
$N$ spins constitute a rectangular cluster.
According to the scaling hypothesis,
the intersection point indicates,
a location of the critical point;
for
details of the scaling theory,
we refer readers to a comprehensive review article \cite{Pelissetto02}.
We see that
a phase transition occurs at $\Gamma \approx 4.8$.

In Fig. \ref{figure4},
we plot the approximate critical point
$\Gamma_c(L_1,L_2)$ for 
$[2/(L_1+L_2)]^3$ with
$12 \le N_1 < N_2 \le 32$
($L_{1,2}=\sqrt{N_{1,2}}$).
[The abscissa scale $1/L^3$ is based on the scaling theory,
which states that
in principle, the corrections of $\Gamma_c$ should behave like
$1/L^{1/\nu+\omega}$ with $1/\nu+\omega=1.5868(3)+0.821(5)$ \cite{Deng03}.
There may also exist subdominant corrections, and
the abscissa scale is set to $1/L^3$.]
Here,
the symbol ($+$) denotes the simulation result
with the optimal
screw pitch adjusted to
Eq. (\ref{adjusted_v}), whereas 
the data of 
($\Box$) were calculated
 with the screw pitch fixed  
to Eq. (\ref{fixed_v}) tentatively.
The approximate critical point 
satisfies the fixed-point relation 
with respect to $L \Delta E$
\begin{equation}
\label{transition_point}
L_1 \Delta E(L_1) |_{\Gamma=\Gamma_c(L_1,L_2)}  = 
L_2 \Delta E(L_2) |_{\Gamma=\Gamma_c(L_1,L_2)}
  ,
\end{equation}
for a pair of system sizes $(L_1,L_2)$.
The data with the conventional scheme [symbol
($\Box$)] exhibit a pronounced bumpy finite-size behavior,
at least, as compared to those of ($+$).
The bottom of the ($\Box$) data corresponds to $N \approx 5^2$ ($1/5^3=0.008$);
for such a quadratic system size, the data suffer from an irregularity 
\cite{Novotny90}
intrinsic to the screw-boundary condition.
Such irregularities appear to be remedied by 
optimizing $v$ [symbol ($+$)] satisfactorily.
The extrapolated critical point $\Gamma_c|_{L\to\infty}$ is no longer
utilized in the subsequent analyses;
rather, the finite-size critical point $\Gamma_c(L_1,L_2)$ is used.
Hence, 
we do not go into further details
as to the reliability of $\Gamma_c$.

In Fig. \ref{figure5},
we present the 
approximate
correlation-length critical exponent
\begin{equation}
\label{exponent_gap}
\nu(L_1,L_2) =
 \frac{\ln (L_1/L_2)}{
\ln \{
\partial_\Gamma[L_1\Delta E(L_1)]/\partial_\Gamma[L_2\Delta E(L_2)]
\} |_{\Gamma=\Gamma_c(L_1,L_2)}
}
  ,
\end{equation}
for $[2/(L_1+L_2)]^2$ with 
$12 \le N_1 < N_2 \le 32$
($L_{1,2}=\sqrt{N_{1,2}}$).
[The abscissa scale $1/L^2$ is based on the scaling theory,
which states that
in principle, the corrections of critical indices should behave like
$1/L^{\omega}$ with $\omega=0.821(5)$ \cite{Deng03}.
Afterward, we make a slight variation of the abscissa scale
so as to appreciate possible subdominant contributions properly.]
The symbols,
($+$) and ($\Box$),
denote the results with the optimal 
[Eq. (\ref{adjusted_v})]
and fixed
[Eq. (\ref{fixed_v})] screw pitches,
respectively.
Again, the former result
($+$)
appears to exhibit
reduced finite-size corrections as compared to the latter one ($\Box$).

We are able to calculate the critical exponent,
based on
 the Binder parameter 
\begin{equation}
U=1-\frac{\langle M^4 \rangle}{3 \langle M^2 \rangle^2}
 ,
\end{equation}
with the magnetization $M=\sum_{i=1}^{N} \sigma^z_i$
and the ground-state expectation $\langle \dots \rangle$.
In Fig. \ref{figure5},
we present the approximate critical exponent
\begin{equation}
\label{exponent_Binder}
\nu(L_1,L_2) =
 \frac{\ln (L_1/L_2)}{
\ln [
\partial_\Gamma U(L_1)/\partial_\Gamma U(L_2)
] |_{\Gamma=\Gamma_c(L_1,L_2)}
}
             ,
\end{equation}
with 
$12 \le N_1 < N_2 \le 32$ and
the approximate critical point $\Gamma_c (L_1,L_2)$
satisfying the fixed-point relation
\begin{equation}
U(L_1) |_{\Gamma=\Gamma_c(L_1,L_2)}=
U(L_2) |_{\Gamma=\Gamma_c(L_1,L_2)}
   .
\end{equation}
The symbols,
($\times$) and ($\circ$),
denote the results 
calculated with the optimal 
[Eq. (\ref{adjusted_v})]
and fixed
[Eq. (\ref{fixed_v})] screw pitches,
respectively.
Again, the former result
($\times$)
appears to exhibit
reduced finite-size deviations.

Apart from an improvement due to the optimal screw pitch,
in Fig. \ref{figure5},
it would be noteworthy that
the data of the symbols, $(+)$ and $(\times)$,
are complementary.
That is,
these results,
obtained independently via
the scaled energy gap 
[Eq. (\ref{exponent_gap})]
and 
the Binder parameter
[Eq. (\ref{exponent_Binder})],
provide the upper and lower bounds for $\nu$, respectively.
Encouraged by this observation,
we carry out an extrapolation to the thermodynamic limit for
$\nu$.
An application of the least-squares fit to the data ($+$)
yields an estimate
$\nu=0.635(4)$ in the thermodynamic limit $L\to\infty$.
Similarly, the result ($\times$)
yields $\nu=0.619(5)$.
An appreciable systematic deviation seems to exist.
Replacing the abscissa scale of Fig. \ref{figure5}
with $[2/(L_1+L_2)]^{1.8}$,
we arrive at the
estimates 
$\nu=0.628(4)$
and 
$\nu=0.624(6)$ for the data ($+$)
and ($\times$), respectively;
the deviation is now within the error margins.
These results may lead to an estimate
$\nu=0.626(10)$
with the error margin covering both systematic and
insystematic (statistical) errors.
As a comparison, we refer to a recent Monte Carlo simulation
result
$\nu=0.63020(12)$
\cite{Deng03}
for the three-dimensional (classical) Ising model.
This result is consistent with ours within the error margins,
suggesting that 
possible systematic errors are appreciated properly.
(As mentioned below,
an application of the
perfect-action technique
might admit a suppression of the error margin.)

Last, we address a remark.
In order to demonstrate a reliability of the present approach,
a comparison with an existing result would be of use;
however,
the Ising model on the triangular lattice has not been studied very thoroughly.
In this respect, we refer readers to Ref. \cite{Nishiyama06} (see Fig. 4),
where the ordinary Ising model was simulated under the screw-boundary condition with
$v=\sqrt{N}$ ($a=1$);
the critical point appears to be close to the existing result.
An aim of this paper is to reduce the 
irregular finite-size corrections of Fig. 4
in Ref. \cite{Nishiyama06}.

\section{\label{section4}
Summary and discussions}

\begin{figure}
\includegraphics{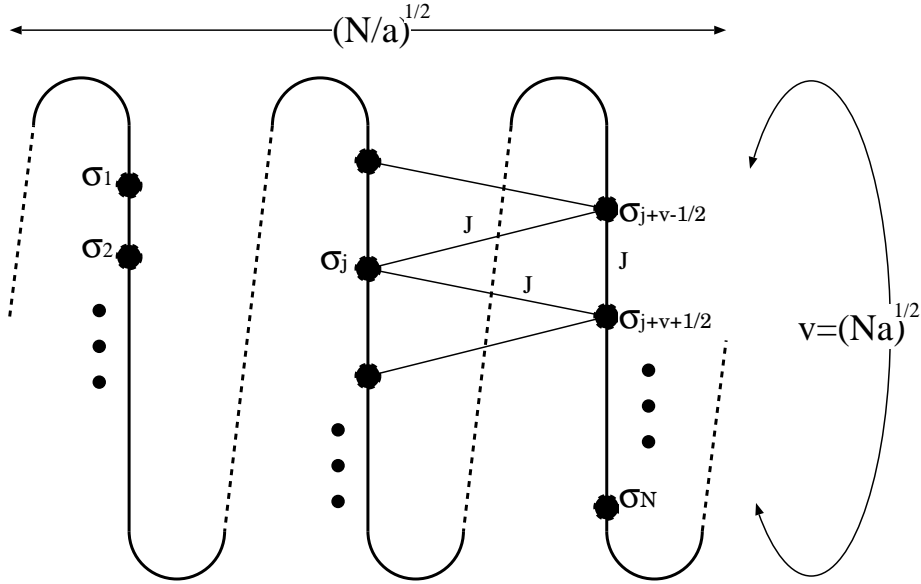}%
\caption{\label{figure1}
A schematic drawing of the spin cluster
for the transverse-field Ising model
(\ref{Hamiltonian})
is presented.
We implement the screw-boundary condition
with the screw pitch $v$, based on Novotny's method.
As indicated above,
the spins
constitute a one-dimensional 
($d=1$)
alignment
$\{ \sigma_i \}$ ($i=1,2,\dots,N$),
and the dimensionality is lifted to 
$d=2$ by the $v(\approx \sqrt{N})$-th-neighbor interactions.
Full details of the simulation scheme are presented in the text.
}
\end{figure}
\begin{figure}
\includegraphics[width=13cm]{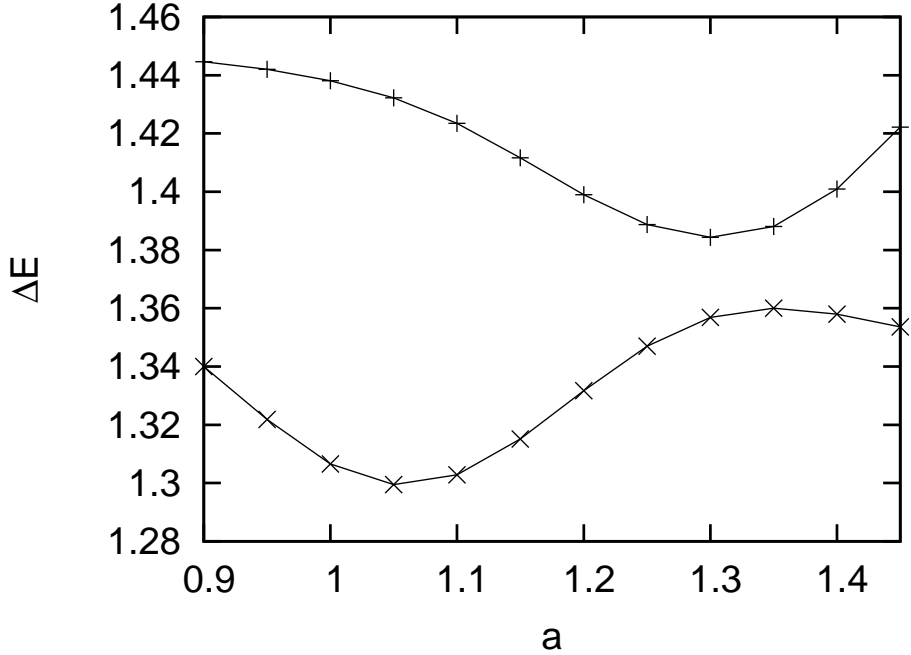}%
\caption{\label{figure2}
The excitation gap $\Delta E$ is presented for a
parameter $a=v^2/N$
($v$ is the screw pitch),
and the system size, $N=$ ($+$) $28$ and ($\times$) $32$;
the transverse magnetic field is fixed to $\Gamma=4.8$,
which is close to a critical point (Fig. \ref{figure3}).
We determine an optimal screw pitch 
$v(N)$
so as to minimize
$\Delta E$ for each $N$.
}
\end{figure}
\begin{figure}
\includegraphics[width=13cm]{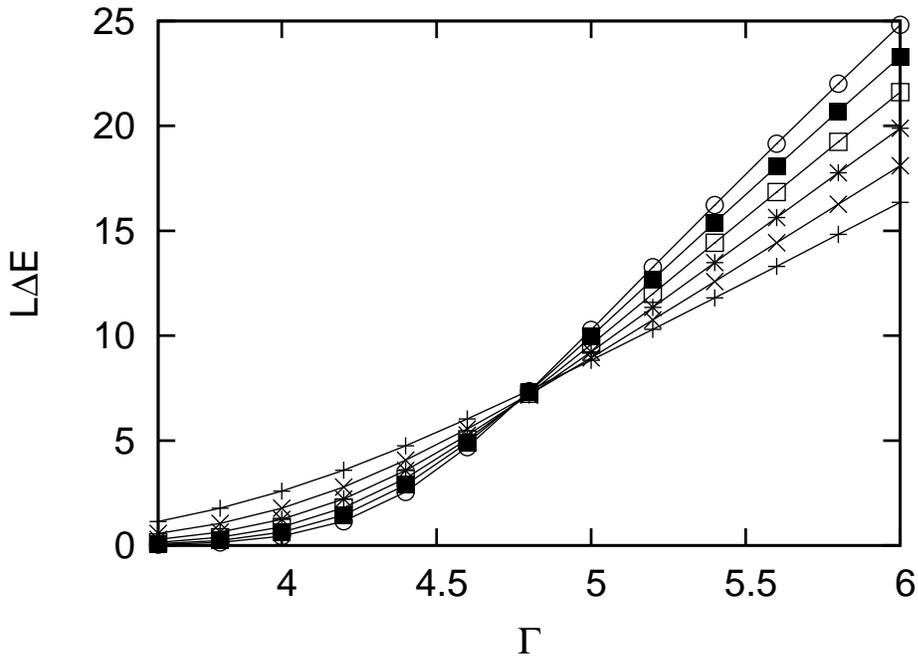}%
\caption{\label{figure3}
The scaled energy gap $L \Delta E$
is plotted for the transverse magnetic field $\Gamma$,
 and the system size 
$N=(+)$ $12$,
($\times$) $16$,
($*$) $20$,
($\Box$) $24$,
($\blacksquare$) $28$, and
($\circ$) $32$ ($L=\sqrt{N}$).
Here, the screw pitch $v$ is set to
an optimal value
[Eq. (\ref{adjusted_v})].
We observe an 
onset of continuous phase transition around $\Gamma \approx 4.8$.
}
\end{figure}
\begin{figure}
\includegraphics{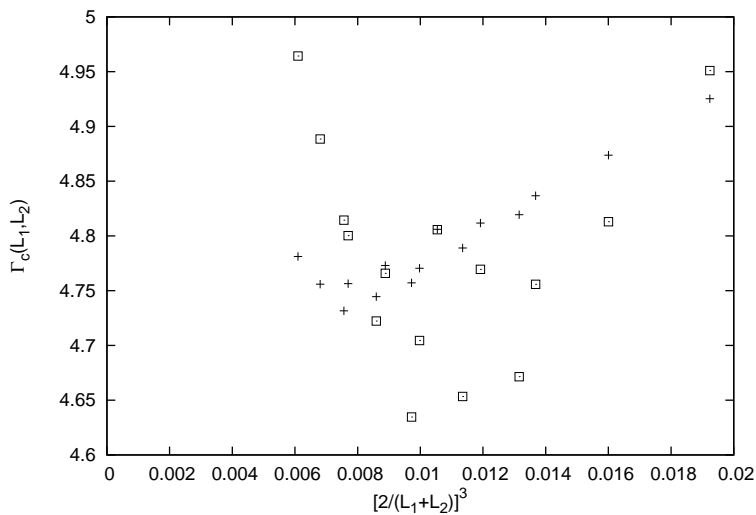}%
\caption{\label{figure4}
The approximate critical point $\Gamma(L_1,L_2)$
[Eq. (\ref{transition_point})]
is plotted for $[2/(L_1+L_2)]^3$ with
$12 \le N_1<N_2 \le 32$ ($L_{1,2}=\sqrt{N_{1,2}}$).
The symbols ($+$) and ($\Box$)
denote the data for the optimal
[Eq. (\ref{adjusted_v})]
and fixed 
[Eq. (\ref{fixed_v})]
screw pitches,
respectively.
The data indicate that the optimal screw pitch
suppresses the (bumpy)
finite-size deviations.
}
\end{figure}
\begin{figure}
\includegraphics[width=13cm]{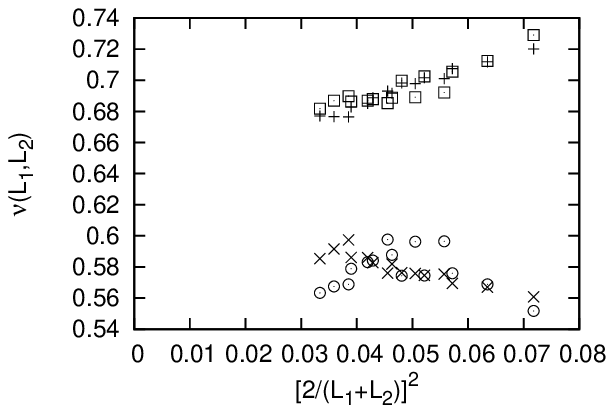}%
\caption{\label{figure5}
The approximate correlation-length critical exponent
$\nu(L_1,L_2)$ is plotted for $[2/(L_1+L_2)]^2$ with
$12 \le N_1<N_2 \le 32$ ($L_{1,2}=\sqrt{N_{1,2}}$).
The data of the symbols, ($+$) and ($\Box$),
were calculated from the scaled energy gap
[Eq. (\ref{exponent_gap})]
for the optimal [Eq. (\ref{adjusted_v})]
and fixed [Eq. (\ref{fixed_v})]
screw pitches, respectively.
Similarly,
from the Binder parameter 
[Eq. (\ref{exponent_Binder})],
 the data of the symbols, ($\times$) and ($\circ$),
were calculated
for the optimal [Eq. (\ref{adjusted_v})]
and fixed [Eq. (\ref{fixed_v})]
screw pitches, respectively.
The data of the optimized screw pitch,
($+$) and ($\times$),
exhibit suppressed finite-size deviations.
An extrapolation to the thermodynamic limit
is argued in the text.
}
\end{figure}

An attempt to suppress the finite-size corrections 
inherent in the screw-boundary condition
(Fig. \ref{figure1})
was made for the transverse-field Ising model
on the triangular lattice.
The screw pitch can afford to vary continuously,
and there is no reason to stick to a fixed pitch $v=\sqrt{N}$.
Setting the transverse magnetic field to a critical-point value
$\Gamma=4.8$,
we scrutinize $v$-dependence of $\Delta E$
(Fig. \ref{figure2}).
Minimizing $\Delta E$,
we determine an optimal value of $v(N)$ for each $N$,
Eq. (\ref{adjusted_v}),
where an effective shape of the cluster
should restore its spatial isotropy.
Actually, the numerical results with the optimal screw pitch
for $N \le 32$
exhibit
suppressed finite-size corrections,
rendering the convergence to the thermodynamic limit
smoother (Figs. \ref{figure4} and \ref{figure5}); 
As a byproduct,
the data obtained from the energy gap
[Eq. (\ref{exponent_gap})]
and the Binder parameter
[Eq. (\ref{exponent_Binder})]
provide the upper and lower bounds for $\nu$,
respectively.
Encouraged by this observation,
we make an extrapolation to $L \to \infty$
to obtain an estimate
$\nu=0.626(10)$ via replacing the abscissa scale with $1/L^{1.8}$.

We mention a number of remarks.
First, as mentioned in Introduction,
the transverse-field Ising model
in $d=2$ and $3$ was reexamined
with the analytical real-space renormalization group
\cite{Miyazaki11};
even for such an analytical tractable system size,
detailed scrutiny of the scaling procedure
affords us
valuable information as to the criticality.
To be specific,
one may eliminate the residual finite-size corrections ($\sim 1/L^{1.8}$)
by 
extending the interaction parameters and tuning them
through the perfect-action technique
\cite{
Chen82,Symanzik83a,
Hasenfratz94,
Blote95,Fernandez94,Ballesteros98}; 
this technique
has been developed in the context of the lattice field theory,
for which the continuum (thermodynamic)
limit has to be undertaken properly and efficiently.
Second,
the present scheme (modification) is readily applicable to the
quantum two-dimensional magnet
\cite{Nishiyama11},
which exhibits a deconfined criticality
(phase transition between the RVB and magnetic phases).
Because of the negative-sign problem, the quantum Monte Carlo method
is not quite efficient to realize the RVB phase,
and
the numerical diagonalization combined with the modified 
screw-boundary condition
would provide a promising candidate so as to survey such an intriguing issue.

\section*{References}
\begin{thebibliography}{10}





\bibitem{Novotny90}M.A. Novotny, J. Appl. Phys. {\bf 67} (1990) 5448.
\bibitem{Novotny92}M.A. Novotny, Phys. Rev. B {\bf 46} (1992) 2939.

\bibitem{Nishiyama11}Y. Nishiyama,
Phys. Rev. B {\bf 83} (2011) 054417.

\bibitem{Singh10}
R.R.P. Singh, Physics {\bf 3} (2010) 35.

\bibitem{Miyazaki11}
R. Miyazaki, H. Nishimori, and G. Ortiz,
Phys. Rev. E {\bf 83} (2011) 051103.


\bibitem{Nishiyama07b}
Y. Nishiyama, Phys. Rev. E {\bf 75} (2007) 051116.

\bibitem{Nishiyama10}Y. Nishiyama,
Nucl. Phys. B {\bf 832} (2010) 605.


\bibitem{Pelissetto02}
A. Pelissetto and E. Vicari,
Phys. Rep. {\bf 368} (2002) 549. 

\bibitem{Deng03}
Youjin Deng and H. W. J. Bl\"ote, Phys. Rev. E {\bf 68} (2003) 036125.


\bibitem{Nishiyama06}
Y. Nishiyama,
Phys. Rev. E {\bf 74} (2006) 016120.



\bibitem{Chen82}J.H. Chen, M.E. Fisher, and B.G. Nickel,
Phys. Rev. Lett. {\bf 48} (1982) 630.
\bibitem{Symanzik83a}K. Symanzik, Nucl. Phys. B {\bf 226} (1983) 187.


\bibitem{Hasenfratz94}P. Hasenfratz and F. Niedermayer,
Nucl. Phys. B {\bf 414} (1994) 785.

\bibitem{Blote95}H.W.J. Bl\"ote, E. Luijten, and J.R. Heringa,
J. Phys. A: Math. Gen. {\bf 28} (1995) 6289.


\bibitem{Fernandez94}L.A. Fern\'andez, A. Mu\~noz Sudupe, J.J. Ruiz-Lorenzo,
and A. Taranc\'on, Phys. Rev. D {\bf 50} (1994) 5935.



\bibitem{Ballesteros98}
H.G. Ballesteros, L.A. Fern\'andez, V. Mart\'in-Mayor, and
A. Mu\~noz Sudupe,
Phys. Lett. B {\bf 441} (1998) 330.




\end{thebibliography}
\end{document}